\documentclass{wsmod}

\newcommand{\rr}{\mbox{\boldmath $r$}}

\newcommand{\rk}{\mbox{\boldmath $k$}}

\newcommand{\rqn}{\mbox{\boldmath $q$}}

\def \pom {{I\!\!P}}

\newcommand{\be}{\begin{equation}}
\newcommand{\ee}{\end{equation}}
\newcommand{\beeq}{\begin{eqnarray}}
\newcommand{\eeeq}{\end{eqnarray}}

\def\lappeq{\mathrel{\rlap{\raise.5ex\hbox{$<$}}
{\lower.5ex\hbox{$\sim$}}}}

\begin{document}

\markboth{V. P. Gon\c{c}alves and M. V. T. Machado}
{Vector Meson Production in Ultraperipheral Heavy Ion Collisions}

%
\catchline{}{}{}{}{}
%

\title{Vector Meson Production in Ultraperipheral Heavy Ion Collisions}

\author{V.P. Gon\c{c}alves}

\address{High and Medium Energy Group (GAME) \\
Instituto de F\'{\i}sica e Matem\'atica, Universidade Federal de
Pelotas\\
Caixa Postal 354, CEP 96010-900, Pelotas, RS, Brazil
\\
barros@ufpel.edu.br}

\author{M.V.T. Machado}

\address{Universidade Estadual do Rio Grande do Sul - UERGS \\
 Unidade Bento Gon\c{c}alves, CEP 95700-000, Bento Gon\c{c}alves, RS, Brazil\\
magno-machado@uergs.edu.br
}

\maketitle


\begin{abstract}
The ultraperipheral heavy ion collisions (UPC's) are an important alternative to study the QCD dynamics until the next generation of 
$e^+e^-/ ep / eA $ colliders become reality. Due to the coherent action of all the protons in the
nucleus, the electromagnetic field is very strong and the
resulting flux of equivalent photons is large, which allows to study two-photon as well as photonuclear interactions at high energies. 
In this paper we present a brief review of the vector meson production in UPC's at high energies using the QCD color dipole approach to describe their photonuclear production and the perturbative QCD Pomeron (BFKL dynamics) to describe the double meson production in photon-photon process. Predictions for rates and integrated cross sections are presented for energies of RHIC and LHC.

\end{abstract}

\ccode{PACS Nos.: 12.38.-t; 24.85.+p 25.75.-q; 14.40.-n  }

\section{Introduction}
In ultraperipheral relativistic heavy-ion collisions (UPC's) the ions do
not interact directly with each other and move essentially
undisturbed along the beam direction. The only possible
interaction is due to the long range electromagnetic interaction
and diffractive processes (For a review see, e. g. Refs.
\cite{bert,bert2}). Due to the coherent action of all the protons in the
nucleus, the electromagnetic field is very strong and the
resulting flux of equivalent photons is large. A photon stemming
from the electromagnetic field of one of the two colliding nuclei
can penetrate into the other nucleus and interact with one or more
of its hadrons, giving rise to photon-nucleus collisions to an
energy region hitherto unexplored experimentally. For example, the
interaction of quasi-real photons with protons has been studied
extensively at the electron-proton collider at HERA, with
 $W_{\gamma p} \le 200$ GeV.
Due to the larger number of photons coming from one of the
colliding nuclei in heavy ion collisions, a  similar and more detailed
study will be possible in these collisions, with $W_{\gamma N}$
reaching 950 GeV for the Large Hadron Collider (LHC) operating in
its heavy ion mode. Similarly, 
one can  investigates QCD effects in the  context of
 coherent two-photon scattering at UPC's \cite{per1,per2,double_meson}.  The important advantage of investigating two-photon
 interactions in nuclear collisions is the flux of virtual photons
 from the electromagnetic field of the nuclei scaling as $Z^2$, giving
 a two-photon luminosity scaling as $Z^4$.  The maximum 
$\gamma \gamma$ collision energy  $W_{\gamma \gamma}$ is $2\gamma 
/R_A$,  about 6 GeV at RHIC and 200 GeV at LHC \cite{bert}. In
 particular, the LHC will have a significant 
energy and luminosity reaching  beyond LEP2, and could be a bridge to
$\gamma \gamma$ collisions at a future $e^+ e^-$ linear collider.

Over the past few  years a comprehensive analysis of the heavy quark \cite{per3,vicber,klein_vogt} and vector meson \cite{per1,per4,double_meson,vicber,klein_nis_prc,strikman,klein_nis_prl} production in UPC's has been made considering different theoretical approaches. For a recent review on heavy quark production in UPC's, see Ref. \cite{mplarevhq}. In particular, much effort has been devoted to obtain signatures of the QCD Pomeron in such processes \cite{per1,per2,per3}, which can be used to constrain the QCD dynamics at high energies. On the other hand, recently the STAR Collaboration released the first data on the cross section of the coherent $\rho$ production in gold - gold UPC's at $\sqrt{s} = 130$ GeV \cite{star_data}, which provides the first opportunity to check the basic features and main approximations of the distinct approaches describing nuclear vector meson photoproduction.
 
Here, we analyze the possibility of
 using UPC's as a photon-photon/nucleus
 collider, and study the production of vector mesons considering different QCD dynamics.  Probably, only
 the next generation of $e^+e^-/ ep / eA $ colliders will allow  to
 discriminate between the distinct approaches for the QCD dynamics at
 high energies. However, until these colliders become reality we need
 to consider alternative searches in the current and/or scheduled  accelerators which allow us to constraint the QCD dynamics. The first analysis, summarized in Sec. \ref{sec2}, concerns to the photonuclear meson production, where we consider the QCD color dipole approach to describe its production in photon-nucleus process. That approach is suitable at high energies and allows take into account the corrections of partons saturation phenomenon (For recent reviews see, e.g. Ref. \cite{hdqcd}) and nuclear effects in a simple and intuitive way. The basic quantity is the dipole cross section, which encodes all information on the interaction of the color dipoles with the nucleus target. The scattering of dipoles off nuclei is modeled through the Glauber-Gribov formalism. The conversion of dipoles on mesons is accounted by the meson wavefunction. Predictions are given for the ion modes and energies of RHIC and LHC. A comparison with the recent STAR data on coherent $\rho$ production is also presented. The second analysis, summarized in Sec. \ref{sect3}, treats the double meson production in two-photon processes, in particular double $J/\Psi$, $\rho-J/\Psi$ and double $\rho$ production. In the case of heavy mesons, we consider the underlying dynamics at high energies as given by the perturbative QCD Pomeron, with the
evolution described by the BFKL equation \cite{bfkl}. Then we investigate the energy behavior in LO and NLO accuracy. The hard scale for the process is guarantee by the large mass of the heavy meson. For the mixed light-heavy meson case, we use the QCD double logarithmic approximation, which depends on the gluon content of the light meson.  The sensitivity to different gluon distributions is addressed. As a byproduct, we obtain the double $\rho$ cross section using the previous results through the Pomeron-exchange factorization theorem. In the last section we discuss the main features of the theoretical estimates and also address the backgroud processes.

\section{Photonuclear vector meson production at UPC's}
\label{sec2}

In heavy ion  collisions the large number of photons coming from
one of the colliding nuclei  will  allow to study photoproduction,
with the 
photonuclear cross sections  given by the convolution between
the photon flux from one of the nuclei and the cross section for
the scattering photon-nuclei. 
The final  expression
for the production of vector mesons in UPC's is then given by,
\begin{eqnarray}
\sigma_{AA \rightarrow AAV}\,\left(\sqrt{S_{\mathrm{NN}}}\right) = \int \limits_{\omega_{min}}^{\infty} d\omega \, \frac{dN\,(\omega)}{d\omega}\,\, \sigma_{\gamma \,A \rightarrow VA} \left(W_{\gamma A}^2=2\,\omega\sqrt{S_{\mathrm{NN}}}\right)\,
\label{sigAA}
\end{eqnarray}
where $\omega$ is the photon energy ($\omega_{min}=m_V^2/4\gamma_L m_p$), $m_V$ is the meson mass and
$\sqrt{S_{\mathrm{NN}}}$ is  the ion-ion c.m.s energy. For instance, the Lorentz factor for LHC is
$\gamma_L=2930$, giving the maximum c.m.s. $\gamma N$ energy
$W_{\gamma A} \lappeq 950$ GeV.  In this process we have that the nuclei are not disrupted and the final state consists solely of the two nuclei and the vector meson decay products. Consequently, we have that the final state is  characterized by a small number of centrally produced particles, with rapidity gaps separating the central final state from both beams. Moreover, due to the coherence requirement, the transverse momentum  is limited to be smaller than $p_T = \sqrt{2}/R_A$, where $R_A$ is the nuclear radius. Therefore, these reactions can be studied experimentally by selecting events with low multiplicity and small total $p_T$.

The photon flux is given by the
Weizsacker-Williams method \cite{bert}. The flux from a charge
$Z$ nucleus a distance $b$ away is
\begin{eqnarray}
\frac{d^3N\,(\omega,\,b^2)}{d\omega\,d^2b}= \frac{Z^2\alpha_{em}\eta^2}{\pi^2 \,\omega\, b^2}\, \left[K_1^2\,(\eta) + \frac{1}{\gamma_L^2}\,K_0^2\,(\eta) \right] \,
\label{fluxunint}
\end{eqnarray}
where $\gamma_L$ is the Lorentz boost  of a single beam and $\eta
= \omega b/\gamma_L$; $K_{0,\,1}(\eta)$ are the
modified Bessel functions. The requirement that  photoproduction
is not accompanied by hadronic interaction (ultraperipheral
collision) can be done by restricting the impact parameter $b$  to
be larger than twice the nuclear radius, $R_A=1.2 \,A^{1/3}$ fm.
Therefore, the total photon flux interacting with the target
nucleus is given by Eq. (\ref{fluxunint}) integrated over the
transverse area of the target for all impact parameters subject to
the constraint that the two nuclei do not interact hadronically.
An analytic approximation for $AA$ collisions can be obtained
using as integration limit $b>2\,R_A$, producing
\begin{eqnarray}
\frac{dN\,(\omega)}{d\omega}= \frac{2\,Z^2\alpha_{em}}{\pi\,\omega}\, \left[\bar{\eta}\,K_0\,(\bar{\eta})\, K_1\,(\bar{\eta})+ \frac{\bar{\eta}^2}{2}\,\left(K_1^2\,(\bar{\eta})-  K_0^2\,(\bar{\eta}) \right) \right] \,
\label{fluxint}
\end{eqnarray}
where $\bar{\eta}=2\omega\,R_A/\gamma_L$.

Analyzing  Eq. (\ref{sigAA})   we obtain that the main input in the calculations of 
 the vector mesons production cross sections in UPC's is the photonuclear cross section $\sigma\,(\gamma A\rightarrow VA)$ which is given by
\begin{eqnarray}
\sigma\,(\gamma A\rightarrow VA) =  \frac{[{\cal I}m \, {\cal
      A}_{\mathrm{nuc}}(s,\,t=0)]^2}{16\pi}\,\,(1+\beta^{2})\,\int_{t_{min}}^\infty
      dt\, |F(t)|^2 \,,
\label{fotonuclear}
\end{eqnarray}
with $t_{min}=(m_V^2/2\,\omega)^2$. Here, the scattering amplitude for meson production on photon-nuclei collisions at zero momentum transfer is labeled by ${\cal A}_{\mathrm{nuc}}$. The quantity $\beta$ is the ratio between the imaginary and real part of the amplitude. Using the
analytical approximation of the Woods-Saxon distribution as a hard
sphere, with radius $R_A$, convoluted with a Yukawa potential with
range $a=0.7$ fm, we obtain that the nuclear form factor reads
as \cite{klein_nis_prc},
\begin{equation}
F(q=\sqrt{|t|}) = \frac{4\pi\rho_0}{A\,q^3}\,
\left[\sin(qR_A)-qR_A\cos(qR_A)\right]
\,\left[\frac{1}{1+a^2q^2}\right]\,\,, \label{FFN}
\end{equation}
where $\rho_0 = 0.16$ fm$^{-3}$. 

Currently, there are in literature different models for the photonuclear vector meson production cross section which differ basically in the treatment for the photon and for its  interaction with the target.
The main aspect is that  real photons have a complicated nature. In a first approximation, the photon is a point-like particle, although in field theory it may fluctuate also into a fermion pair (See discussions in Refs. \cite{bert,nisius}). In the case where  there is a  photon transition in a colorless antiquark-quark pair, the propagation of this colorless hadronic wave packet in a nuclear medium can be treated either in the hadronic basis as a result of Gribov's inelastic corrections or in QCD in terms of the partonic basis, which are complementary. Lets briefly discuss  these two representations (For a detailed discussion see Ref. \cite{Nikolaev}).  The time scale characterizing the evolution of a $q \overline{q}$ wave packet can be estimated based on the uncertainty principle and Lorentz dilation. The lifetime of 
the photon fluctuation is given by $t_c =\nu/(Q^2 +  m^2_V)$,
where  $\nu$ is the photon energy, $m_V$ is the mass of the fluctuation and $Q^2$ is the photon virtuality. It is usually called coherence time. Using light-cone kinematics we can define the coherence length, which is given by $l_c = t_c$. Moreover, one cannot decide whether a ground state $V$ is produced or the next-excited state $V^{\prime}$, unless the process lasts longer than the inverse mass difference between these states. In the rest frame of the nucleus, this formation time is Lorentz dilated and is given by $t_f = 2 \nu/(m^2_{V^{\prime}} - m^2_V)$.  Similarly, we can define a formation length given by $l_f = t_f$. In the hadronic basis, the same process looks quite different. The incident photon may produce different states on a bound nucleon,  the $V$ meson ground state or an excited state. Those states propagate through the nucleus experiencing multiple-diagonal and off-diagonal diffractive interactions, and eventually the ground state is detected. According the quark-hadron duality, we expect that these two descriptions   to be equivalent. However, as these two approaches have been used assuming different approximations, their comparison may provide a scale for the theoretical uncertainty involved.   
Furthermore, it is important to emphasize that at high photon energy $\nu$, both $l_c$ and $l_f$ greatly exceed the nuclear radius $R_A$, which implies in the partonic basis that the transverse size of the $q\overline{q}$ pair do not change during the interaction with the target. This enables one to introduce the QCD dipole picture \cite{nik}, where the process is factorized into the photon fluctuation in a $q\overline{q}$ pair and the dipole cross section.  These aspects become the interpretation in the partonic basis more intuitive and straightforward than in the hadronic basis.

Let us consider the
scattering process $\gamma A \rightarrow V A$ in the QCD dipole approach, where $V$ stands for
both light and heavy mesons. The scattering process can be seen
in the target rest frame as a succession in time of three
factorizable subprocesses: i) the photon fluctuates in a
quark-antiquark pair (the dipole), ii) this color dipole interacts with the
target and, iii) the pair converts into vector meson final state.
Using as kinematic variables the $\gamma^* A$ c.m.s. energy
squared $s=W_{\gamma^* A}^2=(p+q)^2$, where $p$ and $q$ are the target and the
photon momenta, respectively, the photon virtuality squared
$Q^2=-q^2$ and the scaling variable $\tilde{x}=(Q^2 + 4 m_f^2)/(W_{\gamma A}^2+Q^2)$, the
corresponding imaginary part of the amplitude at zero momentum
transfer reads as \cite{kope,nem}, 
\begin{eqnarray}
{\cal I}m \, {\cal A}\, (\gamma A \rightarrow VA)  = \sum_{h, \bar{h}}
\int dz\, d^2\rr \,\Psi^\gamma_{h, \bar{h}}(z,\,\rr,\,Q^2)\,\sigma_{dip}^{\mathrm{target}}(\tilde{x},\rr) \, \Psi^{V*}_{h, \bar{h}}(z,\,\rr) \, ,
\label{sigmatot}
\end{eqnarray}
where $\Psi^{\gamma}_{h, \bar{h}}(z,\,\rr)$ and $\Psi^{V}_{h,
  \bar{h}}(z,\,\rr)$  are the light-cone wavefunctions  of the photon
  and vector meson, respectively. The quark and antiquark helicities are labeled by $h$ and $\bar{h}$
  and reference to the meson and photon helicities are implicitly understood. The variable $\rr$ defines the relative transverse
separation of the pair (dipole) and $z$ $(1-z)$ is the
longitudinal momentum fractions of the quark (antiquark). The basic
blocks are the photon wavefunction, $\Psi^{\gamma}$, the  meson
wavefunction, $\Psi_{T,\,L}^{V}$,  and the dipole-target  cross
section, $\sigma_{dip}^{\mathrm{target}}$.

In the dipole formalism, the light-cone
 wavefunctions $\Psi_{h,\bar{h}}(z,\,\rr)$ in the mixed
 representation $(z,\rr)$ are obtained through two dimensional Fourier
 transform of the momentum space light-cone wavefunctions, 
 $\Psi_{h,\bar{h}}(z,\,\rk)$,  which can be completely determined using light cone perturbation theory. On the other hand, for vector mesons, the light-cone wavefunctions are not known
in a systematic way and they are thus obtained through models (For a recent detailed discussion see Ref. \cite{Nikolaev}).  Here, we follows the
analytically simple DGKP approach \cite{dgkp:97}, which  assumes
that the dependencies on $\rr$ and $z$ of the wavefunction are
factorised, with a Gaussian dependence on $\rr$ (For a detailed discussion see Refs. \cite{sandapen,magno_victor_mesons}). The main shortcoming of this approach is that it breaks the rotational invariance between transverse and longitudinally polarized vector mesons \cite{Nikolaev}. However, as it describes reasonably the HERA data for vector meson production, as pointed out in Ref.
\cite{sandapen}, we will use it in our  phenomenological analysis. 
 Finally, the imaginary part of the forward amplitude can be obtained by
 putting the expressions for photon and vector meson (DGKP) wavefunctions into
 Eq. (\ref{sigmatot}). Moreover, summation over the quark/antiquark
 helicities  and an average over the   transverse polarization states
 of the photon should be taken into account. The transverse component (the longitudinal one does not contribute for photoproduction) is  then written as \cite{kope,nem,sandapen,dgkp:97}
\begin{eqnarray}
{\cal I}m \, {\cal A}_{T} (s,\,t=0) & = &   \int
d^{2}\rr \int_{0}^{1} dz \,\alpha_{{\mathrm{em}}}^{1/2} \,f_{V} \,
f_T(z)\,\exp \left[\frac{-\omega_T^{2}\,\rr^{2}}{2}\right] \nonumber \\
& \times & \left\{\frac{\omega_T^{2}\,\varepsilon \,r}{m_{V}}[z^{2} + (1-z)^{2}]
\,K_{1}(\varepsilon r) + \frac{m_{f}^{2}}{m_{V}}K_{0}(\varepsilon r)
\right \}\sigma_{dip}^{\mathrm{target}}(\tilde{x},\rr) \;,
\label{ampT}
\end{eqnarray}
with $\sigma_{dip}^{\mathrm{target}}$ being the  dipole-nucleus cross
section. In the photoproduction case, $\varepsilon= m_f$, where $m_f$ is the quark mass of flavour $f$. The corresponding parameters for the vector mesons wavefunctions ($m_V$, $\omega_T$, $f_V$, etc) are presented in Table 1 of Ref. \cite{magno_victor_mesons}.  Following Ref. \cite{nem} 
 we have estimated contribution from real part for the photoproduction of vector mesons. 
This correction  is about  3\% for light mesons
and it reaches 13\% for $J/\Psi$ at high energies \cite{magno_victor_mesons}. Additionally for heavy mesons we have take into account
the skewedness effects, associated to 
off-forward features of the process (different transverse momenta
of the exchanged gluons in the $t$-channel),  which  are increasingly
important in this case.

In order to obtain the dipole-nucleus cross section we will 
assume the validity of the Glauber-Gribov picture \cite{gribov} which allows to write \cite{zaka,armesto}
\begin{eqnarray}
\sigma_{dip}^{\mathrm{nucleus}} (\tilde{x}, \,\rr^2;\, A)  = 2\,\int d^2b \,
\left\{\, 1- \exp \left[-\frac{1}{2}\,T_A(b)\,\sigma_{dip}^{\mathrm{proton}} (\tilde{x}, \,\rr^2)  \right] \, \right\}\,,
\label{sigmanuc}
\end{eqnarray}
where $b$ is the impact parameter of the center of the dipole
relative to the center of the nucleus and the integrand gives the
total dipole-nucleus cross section for a  fixed impact parameter.
The nuclear profile function is labeled by $T_A(b)$, which will
be obtained from a 3-parameter Fermi distribution for the nuclear
density. The above equation sums up all the
multiple elastic rescattering diagrams of the $q \overline{q}$
pair and is justified for large coherence length, where the
transverse separation $\rr$ of partons in the multiparton Fock state
of the photon becomes as good a conserved quantity as the angular
momentum, {\it i. e.} the size of the pair $\rr$ becomes eigenvalue
of the scattering matrix.
For the dipole-proton cross section we  follow the
quite successful saturation model \cite{golecwus}, which interpolates between the small and large dipole configurations,
providing color transparency behavior, $\sigma_{dip}^{\mathrm{proton}} \sim \rr^2$,
as $\rr \rightarrow 0$ and constant behavior, $\sigma_{dip}^{\mathrm{proton}} \sim
\sigma_0$, at large dipole separations. The parameters of the
model have been obtained from an adjustment to small $x$ HERA
data.   The parameterization for the dipole cross
section takes the eikonal-like form \cite{golecwus},
\begin{eqnarray}
\sigma_{dip}^{\mathrm{proton}} (\tilde{x}, \,\rr^2)  =  \sigma_0 \,
\left[\, 1- \exp \left(-\frac{\,Q_{\mathrm{sat}}^2(\tilde{x})\,\rr^2}{4} \right) \, \right]\,, \hspace{1.5cm} Q_{\mathrm{sat}}^2(\tilde{x})  =  \left( \frac{x_0}{\tilde{x}}
\right)^{\lambda} \,\,\mathrm{GeV}^2\,,
\label{gbwdip}
\end{eqnarray}
where the saturation scale $Q_{\mathrm{sat}}^2$ defines the onset of the
saturation phenomenon, which depends on energy.  An
additional parameter is the effective light quark mass, $m_f=0.14$
GeV, which plays the role of a regulator for the photoproduction
($Q^2=0$) cross section.  The charm quark mass is
considered to be $m_c=1.5$ GeV. 

\begin{table}[t]
\tbl{\it  The integrated cross section for vector mesons photonuclear production  at UPC's at RHIC and LHC energies. }
{\begin{tabular} {||c|c|c|c|c|c||}
\hline
\hline
& {\bf HEAVY ION}   & $J/\Psi\,(3097)$ & $\phi\,(1019)$ & $\omega\,(782)$ & $\rho\,(770)$  \\
\hline
\hline
 {\bf RHIC} & SiSi &  3.42 $\mu$b & 612 $\mu$b & 764 $\mu$b &  6.74 mb \\
\hline
 & AuAu &  476 $\mu$b &  79 mb & 100 mb & 876 mb \\
\hline
\hline
 {\bf LHC} & CaCa &  436 $\mu$b & 12 mb & 14 mb &  128 mb \\
\hline
&  PbPb &  41.5 mb &  998 mb & 1131 mb & 10069 mb \\
\hline
\hline
\end{tabular}}
\label{tab1}
\end{table}

Let's present the  estimates \cite{per4} from the QCD saturation model for vector meson photoproduction  in the  kinematical range of the colliders RHIC and LHC.
 It is noticeable that  the energy dependence of the dipole-nucleus  cross section [Eq. (\ref{sigmanuc})] 
 is strongly connected with the saturation scale $Q^2_{s\,A}(W_{\gamma\,A}) \propto A^{\frac{1}{3}} \, Q_{\mathrm{sat}}^2$. Namely, the saturation effects are larger  whether the momentum scale is of order or larger than the correspondent size of the vector meson and the energy growth of the cross section is then slowed down.  In  Table \ref{tab1} one shows the corresponding integrated cross section. For LHC energy, one considers lead (Pb) and calcium (Ca), whereas for RHIC one takes gold (Au) and silicon (Si). 
The integrated cross sections can be contrasted with the theoretical estimations using GVDM plus Glauber-Gribov approach of Refs.\cite{strikman} as well as the estimation of Ref. \cite{klein_nis_prc}, which considers VDM plus a classical mechanical calculation for nuclear scattering  and uses as input for the $\gamma \, p \rightarrow V p$ reaction an extrapolation of the experimental DESY-HERA fits for meson photoproduction. Initially lets consider the latter approach (see Tab. III in Ref. \cite{klein_nis_prc}). At RHIC energy and Si nucleus, our results are about 20 \% lower for $\rho$ and $\omega$, whereas gives a larger $\phi$ cross section and almost the same $J/\Psi$ cross section. However, the situation changes for gold nucleus, where the present results are about 50 \% larger than the estimates in Ref. \cite{klein_nis_prc}. At LHC energy and for Ca nucleus, our results gives higher cross section by a factor of order 10 \%, whereas for lead nucleus the factor reaches a factor 2 for light mesons and almost similar for the $J/\Psi$ meson. Basically,  the values are quite similar for light nuclei. For heaviest nuclei, the results overestimate those ones in Ref. \cite{klein_nis_prc} when one considers light mesons and become similar for the $J/\Psi$ case. These features can be understood through the theoretical procedure when considering the nuclear scattering. 

Now we compare our results \cite{per4} with those ones in Ref. \cite{strikman}, where the main focus is on the $\rho$ and $J/\Psi$ production. For $\rho$ the predictions are computed only for RHIC energy $\sqrt{s_{NN}}=130$ GeV and we will consider it later on. We can anticipate that their results are closer to ours since a Glauber-Gribov approach is used in describing the scattering on nuclei. For $J/\Psi$ the theoretical approach for the photonuclear production was the collinear QCD double logarithmic approximation, where the $\gamma A \rightarrow J/\Psi A$ cross section is directly proportional to the squared nuclear gluon density distribution \cite{brodsky}. There, it was considered an impulse approximation (no nuclear shadowing) and a leading twist shadowing version. The impulse approximation gives a larger cross section at central rapidity (about a factor 4  higher for Ca and a factor 6 for Pb), while at fragmentation region both approximations match each other at LHC energy for Ca and Pb nuclei. Our results are closer to their impulse approximation, which suggests nuclear shadowing could be weak for $J/\Psi$ production. This feature can be easily tested in the first experimental measurements of coherent $J/\Psi$ production on UPC's at LHC. Concerning the integrated cross section, they found 0.6 mb for Ca nucleus and 70 mb for Pb at LHC. Our results are 0.44 mb and 41.5 mb, respectively. Thus, our results are about 15 \% lower for Ca and also 40 \% lower for Pb. The difference between the predictions comes from mostly from the distinct QCD approaches considered  used and the  different photon flux in the UPC calculation.

\begin{figure}[t]
\centerline{\psfig{file=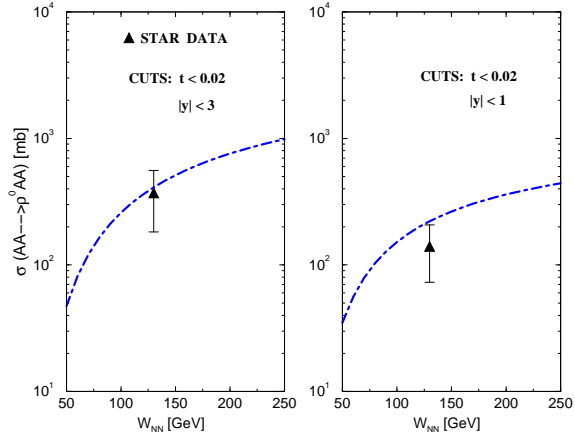,width=75mm}} 
 \caption{\it Energy dependence of coherent $\rho$ meson  production in gold-gold in UPC's at RHIC ($\sqrt{s_{NN}}=130$ GeV). Experimental data from STAR Collaboration.}
\label{fig1}
\end{figure}

Recently, the STAR Collaboration published  the first measuremente for coherent $\rho$ production in gold - gold UPC's at $\sqrt{s} = 130$ GeV \cite{star_data}, providing the first opportunity to test the distinct approaches describing nuclear vector meson photoproduction. Our predictions are presented in Fig. \ref{fig1}. As the  parameters in our approach were  constrained in the study of the photoproduction of vector mesons in $\gamma \,p\, (A)$ interactions \cite{sandapen,magno_victor_mesons}, our predictions are parameter free. In the range of rapidities $|y|\leq 3$, at energy $\sqrt{s_{NN}}=130$ GeV, we have found $\sigma_{\mathrm{sat}}(-3\leq y \leq 3)= 410$ mb in good agreement with the STAR measurement $\sigma_{\mathrm{STAR}}(-3\leq y \leq 3)= 370\pm 170 \pm 80$ mb. For the cut  $|y|\leq 1$, we have obtained $\sigma_{\mathrm{sat}}(-1\leq y \leq 1)= 221$ mb, whereas the STAR result is $\sigma_{\mathrm{STAR}}(-1\leq y \leq 1)= 140\pm 60 \pm 30$ mb.  The values presented here are somewhat similar to the ones obtained in Ref \cite{strikman}, which uses the generalized vector dominance model (GVDM) and the Glauber-Gribov approach, including in addition the finite coherence length effects.

\section{Vector meson production in two-photon processes at UPC's}
\label{sect3}

Relativistic heavy-ion collisions are a potentially prolific
source of $\gamma \gamma$ collisions at high energy colliders. The
advantage of using heavy ions is that the cross sections varies as
$Z^4 \alpha^4$ rather just as $\alpha^4$. Moreover, the maximum
$\gamma \gamma$ collision energy  $W_{\gamma \gamma}$ is $2\gamma
/R_A$,  about 6 GeV at RHIC and 200 GeV at LHC, where $R_A$ is the
nuclear radius and $\gamma$ is the center-of-mass system Lorentz
factor of each ion. In particular, the LHC will have a significant
energy and luminosity reach beyond LEP2, and could be a bridge to
$\gamma \gamma$ collisions at a future $e^+ e^-$ linear collider.
For two-photon collisions, the cross section for the reaction $AA
\rightarrow AA \,V_1 \, V_2$ will be given by
\begin{eqnarray}
\sigma_{AA \rightarrow AA \,V_1 \, V_2} = \int \frac{d
\omega_1}{\omega_1} \, n_1(\omega_1) \int \frac{d
\omega_2}{\omega_2}\, n_2(\omega_2) \,\sigma_{\gamma \gamma
\rightarrow V_1 \, V_2} (W = \sqrt{4 \omega_1 \omega_2}\,)
\,\,,
\end{eqnarray}
where the photon energy distribution $n(\omega)$ is calculated
within the Weizs\"acker-Williams 
approximation \cite{bert}. In general, the  total cross section $
AA \rightarrow AA \,\gamma \gamma \rightarrow AA\, X$, where $X$ is
the system produced within the rapidity gap, factorizes into the
photon-photon luminosity $\frac{d{\cal{L}}_{\gamma
\gamma}}{d\tau}$ and the cross section of the $\gamma \gamma$
interaction,
\begin{eqnarray}
\sigma_{AA \rightarrow AA \,V_1 \, V_2}(s) = \int d\tau \, \frac{d
{\cal{L}}_{\gamma \gamma}}{d\tau} \, \hat \sigma_{\gamma \gamma
\rightarrow V_1 \, V_2}(\hat s), \label{sigfoton}
\end{eqnarray}
where $\tau = {\hat s}/s$, $\hat s = W^2$ is the square of the
center of mass (c.m.s.) system energy of the two photons and $s$
of the ion-ion system. The $\gamma \gamma$ luminosity  is given by
the convolution of the photon fluxes from two ultrarelativistic
nuclei:
\begin{eqnarray}
\frac{d\, {\cal{L}}_{\gamma \gamma}\,(\tau)}{d\tau} = \int ^1 _\tau
\frac{dx}{x} f(x)\, f(\tau/x),
\end{eqnarray}
where the photon distribution function $f(x)$ is related to the
equivalent photon number $n(\omega)$ via $f(x) =
(E/\omega)\,n(xE)$, with $x = \omega/E$ and $E$ is the total
energy of the initial particle in a given reference frame. The
remaining quantity to be determined in order to proceed  is the
quantity $f(x)$, which has been investigated by several groups
(for more details, see e.g. \cite{bert}). Here, we consider
the photon distribution of Ref.\cite{cahn}, providing a photon
distribution which is not factorizable. The authors of \cite{cahn}
produced practical parametric expressions  for the differential
luminosity by adjusting the theoretical results. The comparison with the complete form is consistent within a few percents. The approach given above excludes possible final state
interactions of the produced particles with the colliding
particles, allowing reliable calculations of UPC's. Therefore, in order to estimate the 
double vector meson production cross section in UPC's  it is only necessary to consider a suitable  QCD model
for this process in $\gamma \gamma$ collisions (For a recent analysis see Ref. \cite{Goncalves:2005gv}).

Let's start our dicussion considering the double $J/\Psi$ production. In Ref. \cite{Kwiecinski:1998sa}
 the double $J/\Psi$ production in $\gamma \gamma$ collisions has been
proposed as a probe of the hard QCD Pomeron.
At the photon level we have that the calculation of the cross section
can be made perturbatively, due to the presence of the charm mass.
The non-perturbative content 
is provided only by the $J/\Psi$ light-cone wave function, which
is well constrained through the experimental measurement of its
leptonic width $\Gamma_{J/\Psi\rightarrow \, l^+ l^-}$. In the
following we use the high energy factorization and the BFKL
dynamics in order to perform estimates for the referred reaction.
Assuming a small $t$ approximation,  the total cross section for this process can be written as,
\begin{eqnarray}
\sigma_{tot} \, (\gamma \,\gamma \rightarrow J/\Psi\, J/\Psi) & =
& \frac{1}{B_{J/\Psi\,J/\Psi}}\,\left. \frac{d\sigma \,(\gamma
\,\gamma \rightarrow J/\Psi\, J/\Psi)}{dt}\,\right|_{t=0}\,,
\label{smalltapprox}
\end{eqnarray}
with
\begin{eqnarray}
 \frac{d\sigma \,(\gamma \,\gamma \rightarrow
J/\Psi\, J/\Psi)}{dt} & = &  \frac{|{\cal A}(W^2,t=0)|^2}{16\,\pi}
\, \exp \left(-B_{J/\Psi \, J/\Psi}\cdot |t| \right)\,,
\label{dsdtap}
\end{eqnarray}
where $B_{J/\Psi\, J/\Psi}$ is the corresponding slope parameter, which we assume as being  $B_{J/\Psi\,
J/\Psi}=m_c^{-2}$ (See discussion in Ref. \cite{per1}). In the general case, the 
imaginary part of the scattering amplitude is given as follows \cite{Kwiecinski:1998sa},
\begin{eqnarray}
{\cal I}m\,{\cal A}(W_{\gamma\,\gamma}^2,t) & = &  \int \frac{d^2\rk}{\pi} \,
\frac{\Phi_{\gamma \, J/\Psi}(\rk,\rqn) \,\,
\widetilde{\Phi}_{\gamma \, J/\Psi}(W_{\gamma\,\gamma}^2,\rk,\rqn)}{(\rk + \rqn/2
)^2 \,\,\, (\rk - \rqn/2  )^2}\,,
\end{eqnarray}
where $W$ is the center of mass energy of the two photon system
and the  photon-meson  impact factor is denoted by $\Phi_{\gamma
\, J/\Psi}$.  At the Born level $\widetilde{\Phi} =
\Phi$ and the reaction is described by the two-gluon exchange, which
have transverse momenta $ \rqn/2 \pm \rk$ 
and where the momentum transfer is $t=-\rqn^2$. When considering
the complete gluon ladder contribution, the quantity
$\widetilde{\Phi}$ contains the impact factor and the gluon
emission on the ladder, which is driven by the QCD dynamics. At
the LLA level, the BFKL ladder contribution for the $t$-channel
exchange provides the following expression for it,
\begin{eqnarray}
\widetilde{\Phi}_{\gamma \, J/\Psi}(W_{\gamma\,\gamma}^2,\rk,\rqn) & = & \int
d^2\rk^{\prime} \,\frac{\rk^2}{\rk^{\prime\,2}} \,
{\cal K}(W_{\gamma\,\gamma}^2,\rk,\rk^{\prime},\rqn)\,  \Phi_{\gamma \,
J/\Psi}(\rk^{\prime},\rqn) \,\,,\label{IPBFKL}
\end{eqnarray}
where
\begin{eqnarray}
 \Phi_{\gamma \, J/\Psi}(\rk,\rqn) & = & {\cal C} \, \left[
   \frac{1}{m^2_c + \rqn^2} - 
\frac{1}{m^2_c+ \rk^2} \right]\,\,, \label{IP}
\end{eqnarray}
and ${\cal K}(W_{\gamma\,\gamma}^2,\rk,\rk^{\prime},\rqn)$ is 
solution of the LLA BFKL 
equation at $t\neq 0$. The Eq. (\ref{IP}) defines the impact
factor in the nonrelativistic approximation, where it is assumed that the quark and antiquark have the same longitudinal momentum. A comparison  between the HERA data and the theoretical predictions for $J/\Psi$ photoproduction using this approximation  has found good agreement \cite{FP}.
We have considered
the following parameters for the further calculations:
$m_c=m_{J/\Psi}/2=1.55$ GeV, ${\cal C}= \sqrt{\alpha_{em}}
\,\alpha_s(\mu^2)\, e_c\, \frac{8}{3} \,\pi \, m_c \,f_{J/\Psi}$,
with  $e_c=2/3$ and $ f_{J/\Psi}=0.38$ GeV.

In the Born two-gluon level we have that ${\cal{K}}=\delta^{2}(\rk - \rk^{\prime})$ and the cross section is energy independent. On the other hand,  
in order to perform a LLA BFKL calculation, the following solution
for the evolution equation in the forward case was considered
\cite{per1},  
\begin{eqnarray}
{\cal K}(W_{\gamma\,\gamma}^2,\rk, \rk^{\prime}) & = &
\frac{1}{\sqrt{2\,\pi^3\, a \,\rk^2 \,\rk^{\prime\,2}}}\, 
\frac{1}{\sqrt{\ln (W_{\gamma\,\gamma}^2/\tilde{s})}} \,\left(
  \frac{W_{\gamma\,\gamma}^2}{\tilde{s}} \right)^{\omega_{\pom}} \,
e^{-\frac{\ln^2 (\rk^2/\rk^{\prime\,2})}{2\,a\,\ln
    (W_{\gamma\,\gamma}^2/\tilde{s})} } \,\,, 
\end{eqnarray}
where $ \omega_{\pom}  =  \frac{3\,\alpha_s}{\pi}
\, 4\, \ln \,2$ and $ a=  \frac{3\,\alpha_s}{\pi}
\,28\,\zeta (3)$. The Pomeron intercept is given by $1+\omega_{\pom}$
in the leading 
logarithmic approximation,  which depends on $\alpha_s$. We have
taken $\tilde{s}=1$ GeV$^2$. The results are sensitive to the
choice for the intercept, providing an enhancement of the total
cross section by one/two
 orders of magnitude in relation to the Born level at the
 considered energy range.  
The behavior presented by the approximation above  is
considerably steep on energy. Consequently,  it is important to evaluate the 
corrections which comes from the next-to-leading order corrections to the  BFKL equation or those associated to  unitarity effects.  In order to simulate the
NLO effects, we have used a value $\omega_{\pom}=0.37$.  The effective behavior
obtained is given by $W_{\gamma\,\gamma}^{4\lambda}$, with $\lambda=0.29$, which
value is somewhat close to $\lambda \simeq 0.23-0.29$ obtained in
\cite{Kwiecinski:1998sa}. A more detailed analysis of the double $J/\Psi$  production at the photon level
 can be found in Ref. \cite{per1}.

Having determined the $\gamma \gamma$ cross section we can
estimate the double $J/\Psi$ cross section in UPC's.   In
Ref. \cite{per1} the  energy dependence of the cross
section was investigated,  considering 
the Pb + Pb collisions and different  approximations for the $\gamma
\gamma$ subprocess. Shortly, the LLA BFKL solution
predicts large values of cross section for LHC energies in
comparison with the Born approximation \cite{per1}. The
ratio between the BFKL prediction and the two-gluon cross 
section is around 40 at $\sqrt{s} = 5500$ GeV. Taking into account  non-leading
corrections, this
ratio is reduced to approximately 1.7. Therefore, these results
indicate that if the NLO corrections to the BFKL approach are
account in the approximation considered here, a future
experimental analyzes of double $J/\Psi$ production in
ultraperipheral heavy ion collisions could not easily constrain the QCD
dynamics, since the BFKL(NLO) result is similar to the Born
prediction. Only accurate measurements would allow discriminate
between the two cases. Moreover, it is important to emphasize that
the NLO corrections can be larger, implying a full suppression of
energy enhancement associated with the iteration of gluons in the
$t$-channel  present in the QCD dynamics at high energies.
However, if the energy dependence of the $\gamma \gamma$ cross
section was driven for a large intercept, closer to the LO
prediction, the analysis of UPC's can be useful. It is worth to mention that our results are quite similar 
to those ones computed for the FELIX proposal at LHC \cite{FELIX}.
For PbPb collisions with energies of center of
mass equal to $\sqrt{s}= 5500 \, A$ GeV, luminosities of
${\cal{L}}_{AA} = 4.2 \times 10^{26} \, cm^{-2}\,s^{-1}$ are considered.
Consequently, during a standard $10^6\,s/$ month heavy ion run at
the LHC, we predict approximately  30, 50 and 1100 events for
Born, BFKL (MOD) and BFKL (LO), respectively
\cite{per1}. It is interesting to 
compare these predictions with the results for double $J/\Psi$
production at LEP2, since the $\gamma \gamma$ center of mass
energies are similar. Considering an integrated luminosity of $500
$ pb$^{-1}$ in  three years and $\sqrt{s}= 175$ GeV, almost 70
events are expected taking into account the non-leading
corrections to the BFKL approach \cite{Kwiecinski:1998sa}. Therefore, we
predict a large number of events in ultraperipheral heavy ion
collisions, allowing future experimental analyses, even if the
acceptance for the $J/\Psi$ detection being low. It should be
stressed that the  LHC probably will operate  in its heavy ion
mode only four weeks per year.

Let's now discuss the $\rho J/\Psi$ production in UPC's \cite{double_meson}. At the photon level, this process  can be
calculated similarly as made for  
the   elastic $J/\Psi$ photoproduction off the proton
\cite{Ryskin:1992ui}. Basically, the differential cross section  
is given by \cite{Motyka:2000fc}
\begin{eqnarray}
\frac{d \sigma \,(\gamma \gamma \rightarrow \rho\,
  J/\Psi)}{dt} = {\cal C}\,\frac{16\,\pi^3
  \,g_{\rho}^2\,[\alpha_s(\mu^2)]^2
  \,\,\Gamma_{ee}^{J/\Psi}}{3\,M_{J/\Psi}^5}\,
[\,xG^{\rho}(x,\mu^2)\,]^2\,\, e^{
  -B_{\rho\,J/\Psi}\,|t|}\,\,, 
\label{sigmat} 
\end{eqnarray}
where ${\cal C}$ denotes a product of corrections factors. The two-photon cms
energy is denoted by $W_{\gamma \gamma}$, where
$x=M_{J/\Psi}^2/W_{\gamma \gamma}^2$ and $M_{J/\Psi}$ is the heavy
meson mass. The mass scale is given by $\mu^2=M_{J/\Psi}^2/4$. In the
small-$t$ approximation, the slope is estimated to 
be $B_{\rho\,J/\Psi}=5.5 \pm 1.0$ GeV$^{-2}$ \cite{Motyka:2000fc}
. The light meson-photon coupling is denoted by $g_{\rho}^2=0.454$ and
the heavy meson decay width into a  lepton pair is
$\Gamma_{ee}$. Moreover, $xG^{\rho}(x,\mu^2)$ is the gluon content of
the light meson, which can be contraint from experimental data for the
cross section. 
Currently, there are different parameterizations for this distribution
in the literature, associated to distinct QCD dynamics. In Ref. \cite{double_meson} we
investigate if an experimental analysis of the $\rho\,J/\Psi$ allows  
 discriminate among distinct parameterizations for the gluon content
 of the light meson.  In particular,
we use  both LO and NLO  GRS parameterization   
 \cite{Gluck:1999ub},  the SaS1D parameterization
 \cite{Schuler:1996fc} and a  phenomenological Regge motivated ansatz
 \cite{Motyka:2000fc}. The latter is given by $x\,G^{\rho}_{\mathrm{Regge}} (x,\mu^2) = x_0\,G^{\rho} (x_0, \mu^2) \,\,\left(\frac{x_0}{x} \right)^{\omega_{\pom}-1}$, with $x_0=0.1$  and the Pomeron intercept is considered as
$\omega_{\pom}=1.25$, in agreement with the effective power in the
HERA data. For the nuclear case we allow for a
higher intercept, in order to simulate a BFKL-like behavior, motivated
by the studies in  double $J/\Psi$ production \cite{per1}. The
normalization, $x_0\,G^{\rho} (x_0, M^2_{J/\Psi}/4 )$, is given by the
NLO-GRS parameterization. The differences  among the parameterizations
are sizeable already at photon level as a consequence of distinct
effective exponent $\lambda$. The GRS LO one presents the steeper
behavior, followed by  SaS1D ($\lambda=0.3038$). The Regge motivated
and GRS NLO parameterizations are more close since the intercept for
the  Regge ansatz is similar to the effective power of GRS NLO
($\lambda \simeq 0.228$).  The results for the  $\rho\,J/\Psi$  production cross sections  at the photon level
can be found in Ref. \cite{double_meson}.

For  $\rho\,J/\Psi$  meson production in UPC's at LHC energy range we have 
that an  upper bound is obtained using the SaS1D parameterization and
the lower one by the Regge ansatz with the lower Pomeron intercept
$\omega_{\pom}=1.25$. At the LHC energies $\sqrt{s}=5500$ GeV, the
cross section takes values between 680 nb and  2.6 mb, showing the
process can be used to discriminate among parameterizations. It should
be noted that the ultraperipheral cross section is dominated by not so
high two-photons energies. Although the deviations among the models
are quite visible at very high energies, in the nuclear cross section
such deviations are less sizeable. This is directly related to the
effective two-photon luminosity, which peaks at smaller $W_{\gamma
  \gamma}$. Therefore, correct estimates should include careful
treatment of the low energy threshold effects. The nuclear cross
section is enhanced in relation to the $e^+e^-$ case, which provides
values of hundreds of pb's in contrast with units of mb in the
peripheral nuclear case. 
Comparing our estimates with the FELIX parameterization \cite{FELIX}, it was found  the values predicted by FELIX
collaboration are ever larger than obtained using the approach
presented here, independently of the value of slope used. As a cross
check, we also calculated the cross section for $\rho J/\Psi$ production
in UPC's using the same  kinematical
cut proposed in Ref.  \cite{FELIX}. In such a situation, the agreement
was much better. 

As a byproduct, using the  Pomeron-exchange
 factorization theorem \cite{Gribov:ga}  we can use the  results for
 double $J/\Psi$ production 
and for  $\rho
J/\Psi$ production to  obtain a rough  estimate of the double $\rho$ production in
ultraperipheral heavy ion collisions. 
In principle, this theorem cannot be applied for BFKL Pomeron, since it is not an isolated Regge pole.  As  pointed out in Ref. \cite{russos} (See also Ref. \cite{niko_photon}), only when the asymptotic freedom is incorporated into the BFKL equation is that the BFKL Pomeron can be expressed in terms of a series of isolated poles in the angular-momentum plane, with the contribution of each isolated pole satisfying the factorization theorem. An open question is how much the factorization is violated in different processes. In what follows we estimate the double $\rho$ cross section assuming the   Pomeron-exchange
 factorization theorem. These predictions should be compared with those obtained in a QCD calculation. Preliminary results presented in Ref. \cite{Goncalves:2005gv} shown that the violation is not large. However, more detailed studies are necessary in order  to   quantify this violation. 
Given the assumption that
single Pomeron exchange dominates, the following relation  among the
total cross sections can be posed
\begin{eqnarray}
\sigma(\gamma \gamma \rightarrow \rho \rho) = \frac{(B_{\rho
    J/\Psi})^2}{B_{\rho \rho} \times B_{J/\Psi J/\Psi}} \times
\frac{\left[\sigma (\gamma \gamma \rightarrow \rho
    J/\Psi)\right]^2}{\sigma(\gamma \gamma \rightarrow J/\Psi
  J/\Psi)}\,\,. 
\label{facsig}
\end{eqnarray}
 Using the factorization theorem for the slopes,  we estimate
$B_{\rho \rho} \approx 12.1 \,$ GeV$^{-2}$.    
In the Table \ref{tab2} we present our predictions for the double
$\rho$ production in  ultraperipheral $AA$ collisions, considering
different scenarios for the QCD dynamics. We can see that there is a
large range of possible values for the cross section, which implies
that future experimental data are essential to constraint the
dynamics as well as the violations of the  factorization theorem. It is important to emphasize that our predictions agree with
those presented in Ref. \cite{FELIX}. 

\begin{table}[t]
\tbl{\it The double $\rho$ production cross sections in UPC's at LHC ($\sqrt{s}=5500$ GeV)
for PbPb. }
{\begin{tabular}{||c|c||}
\hline
\hline
 SCENARIO  & $\sigma(A A \rightarrow A A \rho \rho)$ (nb)\\
\hline
BFKL(LO) + GRS(LO)  & $25 \times 10^3$ \\
\hline
BFKL(LO) + GRS(NLO) &  $4 \times 10^3$  \\
\hline
BFKL(LO) + Regge ($\omega_{\pom} = 1.25$) & 810  \\
\hline
BFKL(MOD) + GRS(LO)  & $125 \times 10^4$ \\
\hline
BFKL(MOD) + GRS(NLO) & $20 \times 10^4$  \\
\hline
BFKL(MOD) + Regge ($\omega_{\pom} = 1.25$) & $405 \times 10^2$  \\
\hline
\hline
\end{tabular}}
\label{tab2}
\end{table}

\section{Summary}
\label{end}

Before presenting a summary of the main results, lets discuss the background processes and experimental separation.
In  Ref. \cite{klein_vogt} a detailed 
analysis of the experimental separation between photoproduction
and  two-photon interactions was presented. There, the authors
have estimated that the two-photon cross sections are at least
1000 smaller than the photoproduction cross section, which makes
the experimental separation between the two interactions very
hard. Our calculations  indicate that the inclusion of the QCD
Pomeron effects implies higher cross sections at two-photon level
and, consequently, larger cross sections in ultraperipheral
collisions. Therefore, the inclusion of these effects  implies that,
in general, the contribution of two-photon interactions is
non-negligible. However, the experimental separation 
of two-photon process  still remains a challenge.
 In principle,   
an analysis  of the impact parameter dependence should allow to
separate between the two 
 classes of reactions, since two-photon interactions can occur at
 a significant distance from both nuclei, while a photonuclear
 interaction must occur inside or very  near a nucleus. 

Other possible background process is that associated to
Pomeron-Pomeron interactions.  It has been verified that such
reactions 
 would be non-negligible for light ions, while they are
 significantly suppressed for heavy ions \cite{Roldao:2000ze}. In
 the particular case of the double heavy meson production this
 contribution deserves more detailed studies, since the current
 treatments rely on the Regge formalism instead of a QCD approach. An
 additional contribution in two photon ultraperipheral collisions is
 the meson production accompanied by mutual Coulomb dissociation. 
 It has been estimated that these reactions increase the total cross section
 for an amount of   $\sim 10\%$ at RHIC/LHC energies
 \cite{Baltz:2002pp}. We disregarded this contribution in the present 
 calculations, since  that it would be smaller than the
  associated theoretical uncertainties. 

Let's consider the experimental feasibility of photonuclear production.  Although the vector meson photoproduction at $AA$ collisions to be a small fraction in comparison to the total hadronic cross sections, the experimental separation of these reaction channels is possible.  As photoproduction is an exclusive reaction, $A+A \rightarrow A+A+V$, the separation of the signal from hadronic background would be relatively clear. Namely, the characteristic features in photoproduction at UPC's are low $p_T$ meson spectra and a double rapidity gap pattern. Moreover, the detection (Roman pots) of the scattered nuclei can be an additional useful feature. In hadroproduction, the spectra on transverse momentum of mesons are often peaked around meson mass, $p_T\approx m_V$. An experimental cut $p_T< 1$ GeV would eliminate most part of the hadronic background. Hence, the rapidity cut would enter as an auxiliary separation mechanism. This procedure is specially powerful, since there will be rapidity gaps on both sides of the produced meson. 

As a summary, the cross sections for the $A+A \rightarrow A+A+V$ ($V = \rho, \omega, \phi, J/\Psi$) process were computed and theoretical estimates for scattering on both  light and heavy nuclei are given for RHIC and LHC energies. The rates are very high, mostly for light mesons and at LHC energies. As an important result, we compare our prediction for the coherent $\rho$ meson production with  RHIC data at $\sqrt{s_{NN}}=130$ GeV. The corresponding results are in good agreement with the experimental results when considered the cuts on momentum transfer and on rapidity. Moreover, we have computed the double vector production in UPC where one considers the perturbative QCD Pomeron, described by the BFKL equation. For the mixed light-heavy meson production, we use the DDLA  pQCD approximation, which depends on the gluon content of the light meson and obtain the double $\rho$ cross section using the previous results through the Pomeron-exchange factorization theorem.

As a final remark, in a similar way the strong electromagnetic fields generated by high-energy protons allow us to study photon - nucleon processes in proton - proton interactions in a large kinematical range. These events can be experimentally studied by selecting events with low multiplicity and small total transverse momentum. We have studied this case for the photoproduction of vetor mesons in proton-(anti)proton in Ref. \cite{per4} and the photoproduction of heavy quarks in Ref. \cite{hqppupc}

\section*{Acknowledgments}
  This work was partially
financed by the Brazilian funding agencies CNPq and FAPERGS.


\vspace*{6pt}

\begin{thebibliography}{0}

\bibitem{bert}
 G. Baur, K. Hencken, D. Trautmann, S. Sadovsky, Y. Kharlov, Phys.
Rep. {\bf 364}, 359 (2002).

\bibitem{bert2}
 C.~A. Bertulani, S.~R.~Klein and J.~Nystrand, Ann. Rev. Nucl. Part. Sci. {\bf 55}, 271 (2005). 


\bibitem{per1}
V.~P.~Goncalves and M.~V.~T.~Machado,
Eur.\ Phys.\ J.\ C {\bf 28}, 71 (2003)


\bibitem{per2}
V.~P.~Goncalves and M.~V.~T.~Machado,
Eur.\ Phys.\ J.\ C {\bf 29}, 37 (2003)



\bibitem{double_meson}
V.~P.~Goncalves and M.~V.~T.~Machado,
Eur.\ Phys.\ J.\ C {\bf 29}, 271 (2003).

\bibitem{per3}
V.~P.~Goncalves and M.~V.~T.~Machado,
Eur.\ Phys.\ J.\ C {\bf 31}, 371 (2003).

\bibitem{per4}
V.~P.~Goncalves and M.~V.~T.~Machado,
Eur.\ Phys.\ J.\ C {\bf 40}, 519 (2005).


\bibitem{vicber}
V.~P.~Goncalves and C.~A.~Bertulani,
Phys.\ Rev.\ C {\bf 65}, 054905 (2002).

\bibitem{klein_vogt}
S. R. Klein, J. Nystrand, R. Vogt, Phys. Rev. C {\bf 66}, 044906
(2002).



\bibitem{klein_nis_prc} S. R. Klein, J. Nystrand,  Phys. Rev. C {\bf 60},
014903 (1999).


\bibitem{strikman}
L.~Frankfurt, M.~Strikman and M.~Zhalov,
Phys.\ Lett.\ B {\bf 540}, 220 (2002); Phys.\ Lett.\ B {\bf 537}, 51 (2002); Phys.\ Rev.\ C {\bf 67}, 034901 (2003).



\bibitem{klein_nis_prl}
S.~R.~Klein and J.~Nystrand,
Phys.\ Rev.\ Lett.\  {\bf 92}, 142003 (2004).


 \bibitem{mplarevhq} V.~P.~Goncalves and M.~V.~T.~Machado,
  Mod.\ Phys.\ Lett.\  {\bf 19}, 2525 (2004).


\bibitem{star_data}
C.~Adler {\it et al.}  [STAR Collaboration],
Phys.\ Rev.\ Lett.\  {\bf 89}, 272302 (2002).


\bibitem{hdqcd}
E.~Iancu and R.~Venugopalan,
arXiv:hep-ph/0303204; 
V.~P.~Goncalves, Braz. J. Phys. {\bf 34}, 1406 (2004).

\bibitem{bfkl}
L. N. Lipatov, Sov. J. Nucl. Phys. {\bf 23}, 338 (1976); E. A.
Kuraev, L. N. Lipatov, V. S. Fadin, JETP {\bf 45}, 1999 (1977); I.
I. Balitskii, L. N. Lipatov, Sov. J. Nucl. Phys. {\bf 28}, 822
(1978).


\bibitem{nisius}
R. Nisius, { Phys. Rep.} {\bf 332}, 165 (2000).

\bibitem{Nikolaev}
N.~N.~Nikolaev,
Comments Nucl.\ Part.\ Phys.\  {\bf 21}, 41 (1992); I.~P.~Ivanov, N.~N.~Nikolaev and A.~A.~Savin,
arXiv:hep-ph/0501034.




\bibitem{nik} N. N. Nikolaev, B. G. Zakharov,  Phys. Lett. B  {\bf 332}, 184 (1994);
{Z. Phys. C} {\bf 64}, 631 (1994).


\bibitem{kope}
  B.~Z.~Kopeliovich, J.~Nemchick, N.~N.~Nikolaev and B.~G.~Zakharov,
  Phys.\ Lett.\ B {\bf 309}, 179 (1993); 
  Phys.\ Lett.\ B {\bf 324}, 469 (1994)


\bibitem{nem}
  J.~Nemchik, N.~N.~Nikolaev and B.~G.~Zakharov,
  Phys.\ Lett.\ B {\bf 341}, 228 (1994); 
  J.~Nemchik, N.~N.~Nikolaev, E.~Predazzi and B.~G.~Zakharov,
  Z.\ Phys.\ C {\bf 75}, 71 (1997)



\bibitem{dgkp:97} 
H.~G.~Dosch, T.~Gousset, G.~Kulzinger and H.~J.~Pirner, Phys. Rev. {\bf D55}, 2602 (1997).


\bibitem{sandapen}
J.~R.~Forshaw, R.~Sandapen and G.~Shaw,
 Phys. Rev. D {\bf 69}, 094013 (2004).


\bibitem{magno_victor_mesons}
V.~P.~Goncalves and M.~V.~T.~Machado,
Eur.\ Phys.\ J.\ C {\bf 38}, 319 (2004).


\bibitem{gribov}
V. N. Gribov, Sov. Phys. JETP {\bf 29}, 483 (1969); Sov. Phys. JETP {\bf 30}, 709 (1970).

\bibitem{zaka}
  N.~N.~Nikolaev and B.~G.~Zakharov,
  Z.\ Phys.\ C {\bf 49}, 607 (1991).


\bibitem{armesto}
N. Armesto, Eur. Phys. J. C {\bf 26}, 35 (2002).


\bibitem{devries}
C. W. De Jager, H. De Vries, C. De Vries, Atom. Data Nucl. Data Tabl. {\bf 14}, 479 (1974).

\bibitem{golecwus}  K. Golec-Biernat, M. W\"usthoff,  Phys. Rev. D  {\bf 59},
014017 (1999);  Phys. Rev. D {\bf 60}, 114023 (1999).


\bibitem{brodsky}
S.~J.~Brodsky, L.~Frankfurt, J.~F.~Gunion, A.~H.~Mueller and M.~Strikman,
Phys.\ Rev.\ D {\bf 50}, 3134 (1994).



\bibitem{cahn}
R.~N.~Cahn and J.~D.~Jackson,
Phys.\ Rev.\ D {\bf 42}, 3690 (1990).


\bibitem{Kwiecinski:1998sa}
  J.~Kwiecinski and L.~Motyka,
  Phys.\ Lett.\ B {\bf 438}, 203 (1998).

\bibitem{Goncalves:2005gv}
  V.~P.~Goncalves and W.~K.~Sauter,
 Eur.\ Phys.\ J.\ C {\bf 44}, 515 (2005).


\bibitem{FP}
  J.~R.~Forshaw and G.~Poludniowski,
  Eur.\ Phys.\ J.\ C {\bf 26}, 411 (2003);  G.~G.~Poludniowski, R.~Enberg, J.~R.~Forshaw and L.~Motyka,
  JHEP {\bf 0312}, 002 (2003)

\bibitem{FELIX} A.~Ageev et al., J. Phys. G: Nucl. Part. Phys.\ {\bf 28}, R117 (2002).

\bibitem{Ryskin:1992ui}
M.~G.~Ryskin,
Z.\ Phys.\ C {\bf 57}, 89 (1993).



\bibitem{Motyka:2000fc}
  L.~Motyka and B.~Ziaja,
  Eur.\ Phys.\ J.\ C {\bf 19}, 709 (2001).

\bibitem{Gluck:1999ub}
  M.~Gluck, E.~Reya and I.~Schienbein,
  Phys.\ Rev.\ D {\bf 60}, 054019 (1999)
  [Erratum-ibid.\ D {\bf 62}, 019902 (2000)].
  

\bibitem{Schuler:1996fc}
  G.~A.~Schuler and T.~Sjostrand,
  Phys.\ Lett.\ B {\bf 376}, 193 (1996).
  

\bibitem{Gribov:ga}
V.~N.~Gribov and I.~Y.~Pomeranchuk,
Sov.\ Phys.\ JETP {\bf 15}, 1168L (1962)
[Zh.\ Eksp.\ Teor.\ Fiz.\  {\bf 42}, 1682 (1962)].


\bibitem{russos}
  V.~S.~Fadin, E.~A.~Kuraev and L.~N.~Lipatov,
  Phys.\ Lett.\ B {\bf 60} (1975) 50;  E.~A.~Kuraev, L.~N.~Lipatov and V.~S.~Fadin,
  Sov.\ Phys.\ JETP {\bf 44} (1976) 443
  [Zh.\ Eksp.\ Teor.\ Fiz.\  {\bf 71} (1976) 840]; E.~A.~Kuraev, L.~N.~Lipatov and V.~S.~Fadin,
  Sov.\ Phys.\ JETP {\bf 45} (1977) 199
  [Zh.\ Eksp.\ Teor.\ Fiz.\  {\bf 72} (1977) 377]; L.~N.~Lipatov,
  Sov.\ Phys.\ JETP {\bf 63}, 904 (1986)
  [Zh.\ Eksp.\ Teor.\ Fiz.\  {\bf 90}, 1536 (1986)].

\bibitem{niko_photon}
  N.~N.~Nikolaev, J.~Speth and V.~R.~Zoller,
  J.\ Exp.\ Theor.\ Phys.\  {\bf 93} (2001) 957
  [Zh.\ Eksp.\ Teor.\ Fiz.\  {\bf 93} (2001) 1104]; Eur.\ Phys.\ J.\ C {\bf 22}, 637 (2002).



\bibitem{Roldao:2000ze}
  C.~G.~Roldao and A.~A.~Natale,
  Phys.\ Rev.\ C {\bf 61}, 064907 (2000).
  

\bibitem{Baltz:2002pp}
  A.~J.~Baltz, S.~R.~Klein and J.~Nystrand,
  Phys.\ Rev.\ Lett.\  {\bf 89}, 012301 (2002).
 


\bibitem{hqppupc}
V.~P.~Goncalves and M.~V.~T.~Machado, Phys. Rev. D {\bf 71}, 014025 (2005).






\end{thebibliography}
\end{document}